# A Survey on Preprocessing Methods for Web Usage Data

V.Chitraa
Lecturer
CMS College of Science and Commerce
Coimbatore, Tamilnadu, India
vchit2003@yahoo.co.in

Dr. Antony Selvdoss Davamani
Reader in Computer Science
NGM College (AUTONOMOUS )
Pollachi, Coimbatore,Tamilnadu, India
selvdoss@yahoo.com

*Abstract*— **World Wide Web is a huge repository of web pages and links. It provides abundance of information for the Internet users. The growth of web is tremendous as approximately one million pages are added daily. Users' accesses are recorded in web logs. Because of the tremendous usage of web, the web log files are growing at a faster rate and the size is becoming huge. Web data mining is the application of data mining techniques in web data. Web Usage Mining applies mining techniques in log data to extract the behavior of users which is used in various applications like personalized services, adaptive web sites, customer profiling, prefetching, creating attractive web sites etc., Web usage mining consists of three phases preprocessing, pattern discovery and pattern analysis. Web log data is usually noisy and ambiguous and preprocessing is an important process before mining. For discovering patterns sessions are to be constructed efficiently. This paper reviews existing work done in the preprocessing stage. A brief overview of various data mining techniques for discovering patterns, and pattern analysis are discussed. Finally a glimpse of various applications of web usage mining is also presented.**

## I. INTRODUCTION

Data mining is defined as the automatic extraction of unknown, useful and understandable patterns from large database. Enormous growth of World Wide Web increases the complexity for users to browse effectively. To increase the performance of web sites better web site design, web server activities are changed as per users' interests. The ability to know the patterns of users' habits and interests helps the operational strategies of enterprises. Various applications like e-commerce, personalization, web site designing, recommender systems are built efficiently by knowing users navigation through web. Web mining is the application of data mining techniques to automatically retrieve, extract and evaluate information for knowledge discovery from web documents and services.

The objects of Web mining are vast, heterogeneous and distributing documents. The logistic structure of Web is a graph structured by documents and hyperlinks, the mining results maybe on Web contents or Web structures. Web mining is divided into three types. They are Web content mining, Web structure mining and Web usage mining.

Web Content Mining deals with the discovery of useful information from the web contents or data or documents or services.

Web Structure Mining mines the structure of hyperlinks within the web itself. Structure represents the graph of the link in a site or between the sites.

Web Usage Mining mines the log data stored in the web server.

### A. Web Usage Mining

Web usage mining also known as web log mining is the application of data mining techniques on large web log repositories to discover useful knowledge about user's behavioral patterns and website usage statistics that can be used for various website design tasks. The main source of data for web usage mining consists of textual logs collected by numerous web servers all around the world. There are four stages in web usage mining.

*Data Collection :* users log data is collected from various sources like serverside, client side, proxy servers and so on.

*Preprocessing :* Performs a series of processing of web log file covering data cleaning, user identification, session identification, path completion and transaction identification.

*Pattern discovery :* Application of various data mining techniques to processed data like statistical analysis, association, clustering, pattern matching and so on.

*Pattern analysis :* once patterns were discovered from web logs, uninteresting rules are filtered out. Analysis is done using knowledge query mechanism such as SQL or data cubes to perform OLAP operations.

All the four stages are depicted through the following figure.

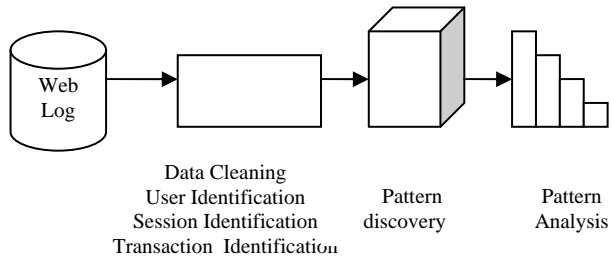

Figure 1. Phases of Web usage mining

The objective of this paper is to provide a review of web usage mining and a survey of preprocessing stage. Data Collection section list out various data sources, preprocessing section reviews the different works done in session identification, path completion process. Remaining sections briefs about pattern discovery, analysis and the different areas of applications where web usage mining is used.

## II. DATA COLLECTION

Data Collection is the first step in web usage mining process. It consists of gathering the relevant web data. Data source can be collected at the server-side, client-side, proxy servers, or obtain from an organization's database, which contains business data or consolidated Web data [13].

*Server level collection* collects client requests and stored in the server as web logs. Web server logs are plain text that is independent from server platform. Most of the web servers follow common log format as

" ipaddress username password date/timestamp url version status-code bytes-sent"

Some servers follow Extended log format along with referrer and user agent. Referrer is the referring link url and user agent is the string describing the type and version of browser software used. Web cache and the IP address misinterpretation are the two drawbacks in the server log. Web cache keeps track of web pages that requests and saves a copy of these pages for a certain period. If there is a request for same page, the cache page is in use instead of making new request to the server. Therefore, these requests are not record into the log files.

*Cookies* are unique ID generated by the web server for individual client browsers and it automatically tracks the site visitors [16]. When the user visits next time the request is send back to the web server along with ID. However if the user wishes for privacy and security, they can disable the browser option for accepting cookies.

*Explicit User Input* data is collected through registration forms and provides important personal and demographic information and preferences. However, this data is not reliable since there are chances of incorrect data or users neglect those sites.

*Client Side Collection* is advantageous than server side since it overcomes both the caching and session identification problems. Browsers are modified to record the browsing behaviors. Remote agents like Java Applets are used to collect user browsing information. Java applets may generate some additional overhead especially when they are loaded for the first time. But users are to be convinced to use modified browser. Along with log files intentional browsing data from client side like "add to my favorites", "copy" is also added for efficient web usage mining [27].

*Proxy level collection* is the data collected from intermediate server between browsers and web servers. Proxy caching is used to reduce the loading time of a Web page experienced by users as well as the network traffic load at the server and client sides [16]. Access log from proxy servers are of same format as web server log and it records the web page request and response for the server. Proxy traces may reveal the actual HTTP requests from multiple clients to multiple Web servers. This may serve as a data source for characterizing the browsing behavior of a group of anonymous users sharing a common proxy server.

## III. DATA PREPROCESSING

The information available in the web is heterogeneous and unstructured. Therefore, the preprocessing phase is a prerequisite for discovering patterns. The goal of preprocessing is to transform the raw click stream data into a set of user profiles [8]. Data preprocessing presents a number of unique challenges which led to a variety of algorithms and heuristic techniques for preprocessing tasks such as merging and cleaning, user and session identification etc [18]. Various research works are carried in this preprocessing area for grouping sessions and transactions, which is used to discover user behavior patterns.

*A. Data Cleaning*

Data Cleaning is a process of removing irrelevant items such as jpeg, gif files or sound files and references due to spider navigations. Improved data quality improves the analysis on it. The Http protocol requires a separate connection for every request from the web server. If a user request to view a particular page along with server log entries graphics and scripts are download in addition to the HTML file. An exception case is Art gallery site where images are more important. Check the Status codes in log entries for successful codes. The status code less than 200 and greater than 299 were removed.

*B. User Identification*

Identification of individual users who access a web site is an important step in web usage mining. Various methods are to be followed for identification of users. The simplest method is to assign different user id to different IP address. But in Proxy servers many users are sharing the same address and same user uses many browsers. An Extended Log Format overcomes this problem by referrer information, and a user agent. If the IP address of a user is same as previous entry and user agent is different then the user is assumed as a new user. If both IP address and user agent are same then referrer URL and site topology is checked. If the requested page is not directly reachable from any of the pages visited by the user, then the user is identified as a new user in the same address [20]. Caching problem can be rectified by assigning a short expiration time to HTML pages enforcing the browser to retrieve every page from the server [7].

C. *Session Identification*

A user session can be defined as a set of pages visited by the same user within the duration of one particular visit to a web-site. A user may have a single or multiple sessions during a period. Once a user was identified, the click stream of each user is portioned into logical clusters. The method of portioning into sessions is called as Sessionization or Session Reconstruction. A transaction is defined as a subset of user session having homogenous pages. There are three methods in session reconstruction. Two methods depend on time and one on navigation in web topology.

*Time Oriented Heuristics :* The simplest methods are time oriented in which one method based on total session time and the other based on single page stay time. The set of pages visited by a specific user at a specific time is called page viewing time. It varies from 25.5 minutes [5] to 24 hours [23] while 30 minutes is the default timeout by R.Cooley [19]. The second method depends on page stay time which is calculated with the difference between two timestamps. If it exceeds 10 minutes then the second entry is assumed as a new session. Time based methods are not reliable because users may involve in some other activities after opening the web page and factors such as busy communication line, loading time of components in web page, content size of web pages are not considered.

*Navigation-Oriented Heuristics :* uses web topology in graph format. It considers webpage connectivity, however it is not necessary to have hyperlink between two consecutive page requests. If a web page is not connected with previously visited page in a session, then it is considered as a different session. Cooley proposed a referrer based heuristics on the basis of navigation in which referrer URL of a page should exists in the same session. If no referrer is found then it is a first page of a new session.

Both the methods are used by many applications. To improve the performance different methods were devised on the basis of Time and Navigation Oriented heuristics by different researchers. Different works were done by researchers for effective reconstruction of sessions.

The referrer-based method and time-oriented heuristics method are combined to accomplish user session identification in [13]. Web Access Log set is the set of all records in the web access log and stored according to time sequence A User Session Set is obtained from the Web Access Log Set by following rules such as different users are distinguished by different IP address. If the IP addresses are same, the different browsers or operating systems indicate different users and if the IP addresses are same, the different browsers and operation systems are same, the referrer information is taken into account. The Referrer URL field is checked and a new user session is identified if the URL in the Referrer URL field has never been accessed before, or there is a large interval (more than 10 seconds [12]) between the access time of this record and the previous one if the Referrer URL field is empty. If the sessions identified by the previous step contain more than one visit by the same user at different time, the time-oriented heuristics is then used to divide the different visits into different user sessions.

A simple algorithm is devised by Baoyao Zhou [4]. An access session is created as a pair of URL and the requested time in a sequence of requests with a timestamp. The duration of an URL is estimated as the difference of request time of successor entry and current entry. For the last URL there is no successor. So the duration is estimated as the average duration of the current session. The end time of session is the start time and duration. This algorithm is suitable when there are more number of URL's in a session. The default time set by author is 30 minutes per session.

Smart Miner is a new method devised by Murat Ali and team [16, 17]. This framework is a part of Web Analytics Software. The sessions constructed by SMART-SRA contains sequential pages accessed from server-side works in two stages and follows Timestamp Ordering Rule and Topology rule. In the first stage the data stream is divided into shorter page sequences called candidate sessions by using session duration time and page stay time rules. In the second stage candidate sessions are divided into maximal sub sessions from sequences generated in the first phase. In the second phase referrer constraints of the topology rule are added by eliminating the need for inserting backward browser moves. The pages without any referrers are determined in the candidate session from the web topology. Then those pages are removed. If a hyperlink exists from the previously constructed session then those pages are appended to the previous sessions. In this sessions are formed one after another. An agent simulator is developed by authors to simulate an actual web user. It randomly generates a typical web site topology and a user agent to accesses the same from its client side and acts like a real user. An important feature of the agent simulator is its ability to model dynamic behaviors of a web agent. Time constraint is also considered as the difference between two consecutive pages is smaller than 10 minutes

Another method using Integer Programming was proposed by Robert F.Dell [21]. The advantage of this method is construction of all sessions simultaneously. He suggests that each web log is considered as a register. Registers from the same IP address and agent as well as linked are grouped to form a session. A binary variable is used and a value of 1 or 0 is assigned depending on whether register is assigned a position in a particular session or not. Constraints such as each register is used at most only once and in only one session for each ordered position. A maximization problem is formulated. To improve the solution time the subset of binary variables is set to zero. An experiment is conducted to show how objective function is varied and results are obtained with raw registers and filtered for MM objects and errors. Unique pages and links between pages are counted. Chunks with same IP address and

*(IJCSIS) International Journal of Computer Science and Information Security,*
*Vol. 7, No. 3, 2010*
agent within a time limit is formed such that no register in one chunk could ever be part of a session in another. Experiment is focused with IP address with high diversity and a higher number of registers. Sessions produced better match an expected empirical distribution.

Graphs are also used for session identification. It gives more accurate results for session identification. Web pages are represented as vertices and hyperlinks are represented as edges in a graph. User navigations are modeled as traversals from which frequent patterns can be discovered. i.e., the sub-traversals that are contained in a large ratio of traversals [22]. A method was proposed by Mehdi Heydari and team [15]. They considered client side data is also important to reconstruct user's session. There are three phases in this method. In first phase an AJAX interface is designed to monitor user's browsing behavior. Events such as Session start, end, on page request, on page load, on page focus are created along with user's interaction and are recorded in session. In the second phase a base graph is constructed using web usage data. Browsing time of web pages in indicated as vertices. Traversal is a sequence of consecutive web pages on a base graph [22]. A database is created with traversals. In phase three graph mining method is applied to the database to discover weighted frequent pattern. Weighted frequent pattern is the pattern when weight of traversal is greater than or equal to a given Minimum Browsing Time.

Another algorithm proposed by Junjie Chen and Wei Liu in which data cleaning and session identification is combined [13] In this deleting the content foreign to mining algorithms gathered from web logs. User activity record is checked, judges whether the record is spider record or not and judges whether it is embedded object in pages or not according to URL of pages requested and site structure graph. Session record is searched if no session exists, a new session is established. If the present session ends or exceeds the preset time threshold, the pattern will ends it and founds a new one. Graph mining methods constructs accurate sessions and the time taken is also comparatively less. More research is to be done in this area.

*D. Path Completion*

There are chances of missing pages after constructing transactions due to proxy servers and caching problems [25][26]. So missing pages are added as follows: The page request is checked whether it is directly linked to the last page or not. If there is no link with last page check the recent history. If the log record is available in recent history then it is clear that "back" button is used for caching until the page has been reached. If the referrer log is not clear, the site topology can be used for the same effect. If many pages are linked to the requested page, the closest page is the source of new request and so that page is added to the session. There are three approaches in this regard.

*Reference Length approach:* This approach is based on the assumption that the amount of time a user spends on a page correlates to whether the page is a auxiliary page or content page for that user. It is expected that the time spent on auxiliary page is small and content page is more. A reference length can be calculated that estimates the cut off between auxiliary and content references. The length of each reference is estimated by taking the difference between the time of the next reference and the current reference. But the last reference has no next reference. So this approach assumes the last one is always a auxiliary reference.

*Maximal Forward Reference:* A transaction is considered as the set of pages from the visited page until there is a backward reference. Forward reference pages are considered as content pages and the path is taken as index pages. A new transaction is considered when a backward reference is made.

*Time Window :* A time window transaction is framed from triplets of ipaddress, user identification, and time length of each webpage up to a limit called time window. If time window is large, each transaction will contain all the page references for each user. Time window method is also used as a merge approach in conjunction with one of the previous methods.

An optimal **a**lgorithm is devised by G.Arumugam and S.Suguna [2] to generate accurate path sequences by using two way hashed structure based access history list to frame a complete path with optimal time. In this tree structure of server pages are searched. There are two problems in this search as backward reference consumes more time in unused pages also and pages which are directly referred from other server's leads to incorrect session identification. To overcome these issues authors gives different algorithm. In this Session Identification algorithm data structures such as Array List to represent Web Logs and User Access List, a Hash table to represent server pages, a two-way hashed structure are utilized. Two way hashed structure is used to store Access History List (AHL) to represent user accessed page sequence. Two hash tables primary and secondary hash tables are used in which primary is used to store sessions and pointers to secondary table which is having a complete path navigation. To solve the time consumption only visited pages are stored in access history list and unused is not considered. Using a single search in history list, the page sequences are directly located. When pages are referred from other servers directly start from the page and not from root. If the page is not available in present sessions, start a new session and we can infer that this is not a backward reference but the page is browsed in another server. This method generates correct complete path than maximal forward and reference length methods.

IV. PATTERN DISCOVERY AND ANALYSIS

Once user transactions have been identified, a variety of data mining techniques are performed for pattern discovery in web usage mining. These methods represent the approaches that often appear in the data mining literature such as discovery of association rules and sequential patterns and clustering and classification etc., [13]. Classification is a supervised learning process, because learning is driven by the assignment of instances to the classes in the training data. Mapping a data item into one of several predefined classes is done. It can be done by using inductive learning algorithms such as decision tree classifiers, naive Bayesian classifiers, Support Vector Machines etc., Association Rule Discovery techniques are

81                    http://sites.google.com/site/ijcsis/
                      ISSN 1947-5500

applied to databases of transactions where each transaction consists of a set of items. By using Apriori algorithm the biggest frequent access item sets from transaction databases that is the user access pattern are discovered. Clustering is a technique to group users exhibiting similar browsing patterns. Such knowledge is especially useful for inferring user demographics in order to perform market segmentation in E-commerce applications or provide personalized web content to pages. Sequential Patterns are used to find inter-session patterns such that the presence of a set of items followed by another item in a time-ordered set of sessions. By using this approach, web marketers can predict future visit patterns which will be helpful in placing advertisements aimed at certain user groups.

Pattern Analysis is the last stage of web usage mining. Mined patterns are not suitable for interpretations and judgments. So it is important to filter out uninteresting rules or patterns from the set found in the pattern discovery phase. In this stage tools are provided to facilitate the transformation of information into knowledge. The exact analysis methodology is usually governed by the application for which Web mining is done. Knowledge query mechanism such as SQL is the most common method of pattern analysis [7]. Another method is to load usage data into a data cube in order to perform OLAP operations.

## V  WEB USAGE MINING APPLICATIONS

Users' behavior is used in different applications such as Personalization, e-commerce, to improve the system and to improve the system design as per their interest etc., Web *personalization* offers many functions such as simple user salutation to more complicate such as content delivery as per users interests. Content delivery is very important since non-expert users are overwhelmed by the quantity of information available online. It is possible to anticipate the user behavior by analyzing the current navigation patterns with patterns which were extracted from past web log. Recommendation systems are the most common application. Personalized sites are example for recommendation systems. *E-Commerce* applications need customer details for Customer Relationship Management. Usage mining techniques are very useful to focus customer attraction, customer retention, cross sales and customer departure. *System Improvement* is done by understanding the web traffic behavior by mining log data so that policies are developed for Web caching, load balancing, network transmission and data distribution. Patterns for detecting intrusion fraud, attempted break-ins are also provided by mining. Performance is improved to satisfy users. *Site Modification* is a process of modifying the web site and improving the quality of design and contents on knowing the interest of users. Pages are re-linked as per customer behavior.

## VI  FUTURE

There are a number of issues in preprocessing of log data. Volume of requests in web log in a single log file is the first challenge. Analyzing web user access log files helps to understand the user behaviors in web structure to improve the design of web components and web applications. Log includes entries of document traversal, file retrieval and unsuccessful web events among many others that are organized according to the date and time. It is important to eliminate the irrelevant data. So cleaning is done to speed up analysis as it reduces the number of records and increases the quality of the results in the analysis stage. Efforts in this data to find accurate sessions are likely to be the most fruitful in the creation of much effective web usage mining and personalization systems. By following data preparation steps, it is very easier to generate rules which identify directories for website improvement. More research can be done in preprocessing stages to clean raw log files, and to identify users and to construct accurate sessions.

## VII  CONCLUSION

Web sites are one of the most important tools for advertisements in international area for universities and other foundation. The quality of a website can be evaluated by analyzing user accesses of the website. To know the quality of a web site user accesses are to be evaluated by web usage mining. The results of mining can be used to improve the website design and increase satisfaction which helps in various applications. Log files are the best source to know user behavior. But the raw log files contains unnecessary details like image access, failed entries etc., which will affect the accuracy of pattern discovery and analysis. So preprocessing stage is an important work in mining to make efficient pattern analysis. To get accurate mining results user's session details are to be known. The survey was performed on a selection of web usage methodologies in preprocessing proposed by research community. More concentration is done on preprocessing stages like session identification and path completion and we have presented various works done by different researchers. Our research in future is to create more efficient session reconstructions through graphs and mining the sessions using graph mining as quality sessions gives more accurate patterns for analysis of users.

### AUTHORS PROFILE

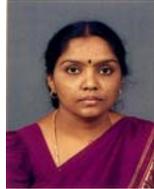

Mrs. V. Chitraa is a doctoral student in Manonmaniam Sundaranar University, Tirunelveli, Tamilnadu. She is working as a lecturer in CMS college of Science and Commerce, Coimbatore. Her research interest lies in Database Concepts, Web Usage Mining, Clustering.

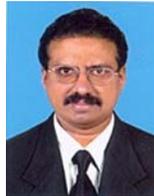

Dr. Antony Selvadoss Davamani is working as a Reader in NGM college with a teaching experience of about 22 years. His research interests includes knowledge management, web mining, networks, mobile computing, telecommunication. He has published about 8 books and 16 papers.